\newcommand{\lya}{\mbox{Ly$\alpha$}}

\newcommand{\hI}{\mbox{H{\sc i}}}

\newcommand{\oVI}{\mbox{O{\sc vi}}}
\newcommand{\cII}{\mbox{C{\sc ii}}}
\newcommand{\cIII}{\mbox{C{\sc iii}}}
\newcommand{\cIV}{\mbox{C{\sc iv}}}

\newcommand{\neVIII}{\mbox{Ne{\sc viii}}}

\newcommand{\ergseccmarcsec}{\mbox{erg~s$^{-1}$~cm$^{-2}$~arcsec$^{-2}$}}

\newcommand{\percc}{\mbox{cm$^{-3}$}}

\newcommand{\nodata}{...}

\documentclass[onecolumn, astrosymb,]{aastex701}
\usepackage{enumitem}
\usepackage{xspace}
\usepackage{amsmath}
\usepackage{comment}
\usepackage{mdframed}
\usepackage[dvipsnames]{xcolor}
\newcommand{\funcol}{RedViolet}

\usepackage{amsmath}
\usepackage{booktabs}
\usepackage{soul}

\graphicspath{{figs/}}
\begin{document}

\title{Astronomical Advantages\footnote{`Astronomical Advantages' pays homage to Lyman Spitzer's 1946 RAND Project report \emph{Astronomical Advantages of an Extraterrestrial Observatory}, which presented the concept of the \emph{`large orbital observatory'} we  hope to boost.  It seems fitting to re-use it here.
} of a  Boost Mission to Facilitate HST Science into the 2030s:\\Imaging the Circumgalactic Medium of Galaxies}

\author[0000-0001-8587-218X]{Matthew J. Hayes}
\affiliation{Stockholm University, Department of Astronomy and Oskar Klein Centre for Cosmoparticle Physics, AlbaNova University Centre, SE-10691, Stockholm, Sweden}
\email[show]{matthew.hayes@astro.su.se}   

\author[0000-0001-6248-1864]{Kate H. R. Rubin}
\affiliation{Department of Astronomy, San Diego State University, San Diego, CA 92182, USA}
\email{krubin@sdsu.edu} 

\author[0000-0002-8518-6638]{Michelle A. Berg}
\affiliation{Department of Physics \& Astronomy, Texas Christian University, Fort Worth, TX 76109, USA}
\email{m.a.berg@tcu.edu}

\author[0000-0002-1002-3674]{Kevin France}
\affiliation{Laboratory for Atmospheric and Space Physics, University of Colorado Boulder, Boulder, CO 80309}
\affiliation{Department of Astrophysical and Planetary Sciences, University of Colorado Boulder, Boulder, CO 80309}
\affiliation{Center for Astrophysics and Space Astronomy, University of Colorado Boulder, Boulder, CO 80309}
\email{xxx}

\author[0000-0002-5808-1320]{Sally Oey}
\affiliation{Department of Astronomy, University of Michigan, Ann Arbor, MI 48109, USA.}
\email{msoey@umich.edu} 

\author[0000-0001-7472-3824]{Ramona Augustin}
\affiliation{Leibniz-Institut f{\"u}r Astrophysik Potsdam (AIP), An der Sternwarte 16, 14482 Potsdam, Germany}
\email{raugustin@aip.de}

\author[0000-0002-1979-2197]{Joseph N. Burchett}
\affiliation{Department of Astronomy, New Mexico State University, PO Box 30001, USA}
\email{xxx} 

\author[0000-0003-4166-2855]{Cody A. Carr}
\affiliation{Department of Astronomy, The University of Michigan, 1085 S. University Avenue, West Hall 323 
Ann Arbor, MI 48109, USA}
\email{codycarr@umich.edu}

\author[0000-0002-2583-5894]{Alison L. Coil}
\affiliation{Department of Astronomy and Astrophysics, University of California, 9500 Gilman Drive, La Jolla, CA 92093, USA}
\email{xxx}

\author[0000-0002-0159-2613]{Sophia R. Flury}
\affiliation{Institute for Astronomy, University of Edinburgh, Royal Observatory, Edinburgh, EH9 3HJ, UK}
\email{xxx}

\author[0000-0002-3817-8133]{Cameron Hummels}
\affiliation{California Institute of Technology, 1200 East California Blvd, Pasadena, CA 91125, USA}
\email{chummels@caltech.edu}

\author[0000-0002-2587-2847]{Varsha P. Kulkarni}
\affiliation{Department of Physics and Astronomy, University of South Carolina, Columbia, SC 29208, USA}
\email{xxx}

\author[0000-0003-0503-4667]{Stephan R. McCandliss}
\affiliation{Johns Hopkins University, Department of Physics \& Astronomy, Center for Astrophysical Sciences, 3400 North Charles Street, Baltimore, MD, USA, 21218}
\email{stephan@pha.jhu.edu}

\author[0000-0003-2589-762X]{Matilde Mingozzi}
\affiliation{AURA for ESA, Space Telescope Science Institute, 3700 San Martin Drive, Baltimore, MD 21218, USA}
\email{mmingozzi@stsci.edu} 

\author[0000-0001-8421-5890]{Dylan Nelson}
\affiliation{Universit\"{a}t Heidelberg, Institut f\"{u}r Theoretische Astrophysik, Zentrum f\"{u}r Astronomie, AU-Str. 2, 69120 Heidelberg, Germany}
\email{dnelson@uni-heidelberg.de} 

\author[orcid=0000-0003-3467-6810]{Zixuan Peng}
\affiliation{Department of Physics, Broida Hall, University of California at Santa Barbara, Santa Barbara, CA 93106, USA}
\email{zixuanpeng@ucsb.edu} 

\author[0000-0001-7016-5220]{Michael Rutkowski}
\affiliation{Department of Astronomy, University of Florida, 211 Bryant Space Science Center, Gainesville, FL 32611, USA}
\email{mjrutkowski.astronomy@gmail.com}

\author[0000-0001-8419-3062]{Alberto Saldana-Lopez}
\affiliation{Stockholm University, Department of Astronomy and Oskar Klein Centre for Cosmoparticle Physics, AlbaNova University Centre, SE-10691, Stockholm, Sweden}
\email{alberto.saldana-lopez@astro.su.se}

\author[0000-0002-7982-412X]{Jason Tumlinson}
\affiliation{Space Telescope Science Institute, Baltimore, MD 21218, USA}
\affiliation{Department of Physics and Astronomy, Johns Hopkins University, Baltimore, MD 21218, USA}
\email{xxx}

\author[0000-0002-7327-565X]{Sarah Tuttle}
\affiliation{Department of Astronomy, University of Washington, Seattle, WA 98195, USA}
\email{xxx} 

\author[0000-0002-6301-638X]{Freeke van de Voort}
\affiliation{Cardiff Hub for for Astrophysics Research and Technology, School of Physics and Astronomy, Cardiff University, Cardiff CF24 3AA, UK}
\email{xxx} 

\author[0000-0002-0507-7096]{Bart P. Wakker}
\affiliation{Eureka Scientific, Inc., 2452 Delmer Street, Oakland, CA 94602, USA}
\email{xxx}

\author[0000-0002-0355-0134]{Jessica K. Werk}
\affiliation{Department of Astronomy, University of Washington, Seattle, WA 98195, USA}
\email{jwerk@uw.edu}


\begin{abstract}
We present the case for imaging ultraviolet line emission from highly ionized metals and \hI\ \lya\ in the circumgalactic medium of galaxies, should the Hubble Space Telescope receive an orbital boost. Hubble can uniquely probe emission lines with ionization potentials between 13 and 200 electron-volts (\lya, \cIV, \oVI, \neVIII, etc). Spatial mapping of the diffuse material traced by these transitions is critical to constraining the physics of feedback and the energetic exchange between galaxies and their circumgalactic environments, as well as basic morphologies of the dominant mass component. Deep high-resolution mapping of these features will not be possible with any other observatory, existing or planned, until HWO is launched, which leaves HST as a critical observatory to test key science drivers for HWO. If HST receives an orbital boost, it can (a) provide the first statistical constraints on the spatial distribution of warm-hot CGM and (b) provide important avenues for science case development, as well as target/pointing selection, for HWO's upcoming spectroscopic facilities. 
\end{abstract}

\section{Introduction -- Importance of the CGM and Major Open Questions}\label{sect:intro}

Galaxy formation, assembly, evolution, and ultimately star formation, are supported by the delivery of gas to galaxy disks via some combination of inflows, mergers, and recycling.  As stars form, they self-regulate the process, providing (mostly) negative feedback from winds and ultraviolet (UV) radiation from massive stars, followed by the explosion of these stars as supernovae (SNe). This process drives galaxy-scale winds and outflows that compete with the inflowing material through hydrodynamical interactions on a battleground we call the circumgalactic medium (CGM, \citealt{Tumlinson.2017, Faucher-Giguere.2023, Crain.2023}).  \emph{What happens to the infalling material, what is the fate of ejected gas, and how is star formation ultimately fueled?}  The answers to these questions are critical for models of galaxy growth, since this material represents the fuel that will/would produce stars on timescales $\sim 1$ Gyr in the future. Due to this gap in our knowledge, exploring cosmic ecosystems was designated as one of the three Priority Areas targeted by the Astro2020 Decadal survey \citep{astro2020}.

Gas that is shock-heated by SN explosions begins its outflow at temperatures almost $10^8$\,K, while inflowing material gets shocked to around the virial temperature of $\sim 10^5 - 10^6$\,K depending on the halo mass of the galaxy (e.g., \citealt{Birnboim.2003,Dekel2009,Correa.2018}). Regardless of the origins we consider, the material filling the CGM must cool substantially if it is to somehow reach the disk,  become the $\lesssim 30$\,K gas that makes up molecular clouds, and ultimately end up inside a star (e.g., \citealt{Maller2004,McCourt.2012,Voit.2015}). Remarkably, observations of both the extended CGM and galaxy outflows identify gas over essentially this full range of temperatures.  Winds, such as those in M82, show gas spanning seven orders of magnitude in temperature (\citealt{Strickland2009} c.f. \citealt{Leroy.2015}), which at the resolution of many of the observations often appear cospatial (although very small scale structures appear through optical line emission; \citealt{Mutchler.2007}).  CGM absorption line studies show features of \cII\ and \oVI\ (factor of 100 in $T$)  that often (but not always) share bulk kinematic properties \citep[e.g.,][]{Werk.2016,Rudie.2019,Wotta.2019}. If gas is to transition from $>10^6$\,K to $<10^4$\,K, it must cool through intermediate temperatures between $10^5$ and $10^6$\,K.  In this regime the negative slope of the cooling function implies an unstable process\textemdash once gas starts cooling it cannot be stopped\textemdash and an almost immediate loss of thermal energy occurs \citep[e.g.,][]{Maller2004}. This cooling is dominated by emission lines of ionized metals. Most of the energy loss is through unobservable UV lines. The main observable line is that of \oVI, which we therefore argue must be observed.

\begin{figure*}[h!]
\noindent
\vspace{-3.5ex} 
\begin{mdframed}[userdefinedwidth=1.0\textwidth, backgroundcolor=\funcol!5!white,linecolor=\funcol!75!black,linewidth=1pt, align=center]
\vspace{-2pt}
\noindent\
\normalsize
The following major questions remain open in our understanding of the conditions in circumgalactic media: 
\begin{enumerate}[topsep=1pt,itemsep=0ex,partopsep=1pt,parsep=1pt]
\item{What is the origin of warm-hot ($10^{5-6}$\,K) material in the CGM? Is it volume filling or does it form in hydrodynamic interactions where hot outrushing gas meets ambient cold material?}

\item{Is the cooler material lifted from the ISM by the wind, or does it form in situ by gas cooling/condensation?}

\item{What is the morphology of the warm gas?  At what radial distances is mechanical energy deposited in extended galaxy halos? (thereby being drained from the wind) }

\item{What are the volume densities and sizes of the gas phases that are seen with absorption line studies? }

\item{How do the conditions in the CGM connect to the properties of the massive stars, and possible AGN activity, that drive the feedback?}

\item{What physical processes drive the cooling and re-accretion of circumgalactic gas?}
\end{enumerate}

The answers to all these questions lie in the combination of UV emission line imaging and absorption spectroscopy of the same species.  Prior to HWO this can only be delivered by a boosted HST. 

\end{mdframed}\vspace{-1mm} 
\end{figure*}

\section{Observing Strategies}\label{sect:observations}

\subsection{Ultraviolet spectroscopy}

The overwhelming majority of what we know of the CGM comes from absorption line spectroscopy, using quasar-galaxy pairs. This approach has a long history, going back to HST's first generation spectrographs of the \emph{Faint Object Spectrograph} \citep[FOS,][]{Bahcall.1996} and the \emph{Goddard High Resolution Spectrograph} \citep[GHRS,][]{Savage.1998,Tripp.1998}, later followed at higher resolution with the 
\emph{Space Telescope Imaging Spectrograph} \citep[STIS,][]{Prochaska.2004,Howk.2009}. More recently the \emph{Cosmic Origins Spectrograph} (COS) has transformed the field with its broad spectral bandpass \citep[e.g.,][]{Tumlinson.2013, Werk.2014, Borthakur.2013, Bordoloi2013, Burchett.2019, Qu.2024}.  Almost invariably, the strategy is to identify UV-bright QSOs behind foreground galaxies; the galaxy redshift range and grating configuration then determines what absorption lines---species, ionization state/temperature regime---are targeted by the survey. 

Thus the CGM is studied in a statistical/ensemble fashion by selecting a range of impact parameters and galaxy properties (e.g. SFR or stellar mass). Physical conditions such as gas temperature, density, and geometric size of clouds (column length) are then inferred by means of photoionization modeling and similar techniques \citep[e.g.,][]{Tumlinson.2011.apj,Peeples.2014,Werk.2014,Qu.2024} under equilibrium assumptions and assessed for the ensemble of different galaxies.  These methods demonstrate that a very significant fraction of a galaxy's baryonic mass may be contained within this warm-hot reservoir near 300,000\,K on the unstable part of the cooling function.  Intriguingly, current hydrodynamical simulations (e.g. FIRE-2/FOGGIE, \citealt{Wijers.2024,Augustin.2025}) are not able to match either the dispersion or the high observed column density of highly ionized gas. These variations in the statistics derived from ensembles are likely driven by small-scale variation \emph{within galaxies}, probably implying small dense clumps and turbulent radiative mixing layers (TRML) that are not captured at tractable spatial resolutions of simulations where the gas phases interact \citep[e.g.,][]{Fielding.2022,Chen.2025}. The paucity of UV-bright QSOs makes the `random' positioning of quasar sightlines an inefficient way to sample (for example) interactions that occur near the disk interface regions from where winds are launched\footnote{at least for commonly observed QSO magnitudes; see the submitted White Paper by S. Borthakur for an alternative.}.  

\begin{figure}[h]
    \centering
    \includegraphics[width=1\linewidth]{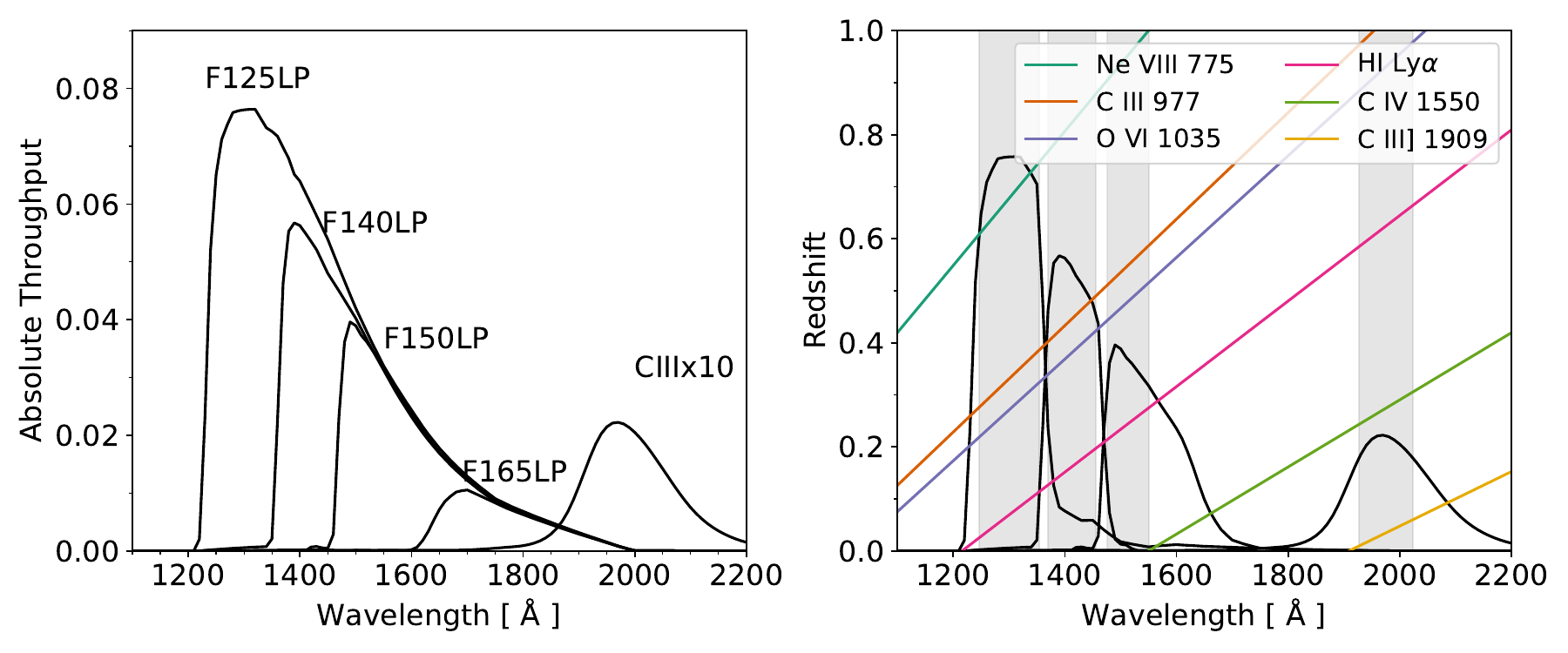}
    \caption{\emph{Left}: Relevant longpass filters in ACS/SBC and the CIII filter in STIS.  \emph{Right}: Synthetic bandpasses obtained by subtraction of adjacent filter pairs. Throughput above 80\% of peak is shaded in gray.  The wavelength of different emission lines as a function of redshift from $z=0-1$ are shown by colored lines.  Possible multi-line combinations are listed in Table~\ref{tab:filter_redshifts}.}
    \label{fig:bandpass}
\end{figure}

\begin{deluxetable}{c l l}
\tablecaption{Filters and Redshifts for Example Observations.}
    \tablehead{Redshift Range & Line combinations (Filters) & Example publications} 
    \startdata
        $0.03 - 0.06$ & \lya\ (F125LP) + \cIV\ (F150LP) + \cIII] (STIS/CIII) & 1,2,3 \hspace{5cm} \\
        $0.22 - 0.30$ & \oVI\ (F125LP) + \lya\ (F150LP) + \cIV\ (STIS/CIII)  & 4,5 \\
        $0.31 - 0.32$ & \cIII\ (F125LP) + \lya\ (F150LP) + \cIV\ (STIS/CIII) & \nodata \\
        $0.33 - 0.38$ & \cIII\ (F125LP) + \oVI\ (F140LP)   & \nodata \\
        $0.43 - 0.49$ & \cIII\ (F140LP) + \oVI\ (F150LP)   & \nodata \\
        $0.57 - 0.59$ & \cIII\ (F150LP) + \lya\ (STIS/CIII) & \nodata \\
        $0.62 - 0.65$ & \neVIII\ (F125LP) + \cIII\ (F150LP) + \lya\ (STIS/CIII) & \nodata \\
        $0.62 - 0.70$ & \neVIII\ (F125LP) + \lya\ (STIS/CIII) & \nodata \\
        $0.83 - 0.86$ & \neVIII\ (F140LP) + \oVI\ (STIS/CIII) & \nodata \\
    \enddata
\label{tab:filter_redshifts}
\tablecomments{Combinations of emission lines that can be imaged with HST.  Listed filters in parentheses transmit the line; continuum must also be sampled by the next filter in the set (see Figure~\ref{fig:bandpass}).   The \cIII]\,$\lambda\lambda 1907$,09\,\AA\ emission line is always denoted with the intercombination specifier `]' to distinguish it from the permitted resonance line at $977$\,\AA.  Individual emission lines can of course be imaged over larger redshift ranges.  Example references are 1=\citep{Hayes.2013}, 2=\citep{Oey.2023}, 3=\citep{Micheva.2020}, 4=\citep{Hayes.2016}, 5=\citep{Ha.2025}. }
\end{deluxetable}

\begin{figure}
    \centering
    \includegraphics[width=1\linewidth]{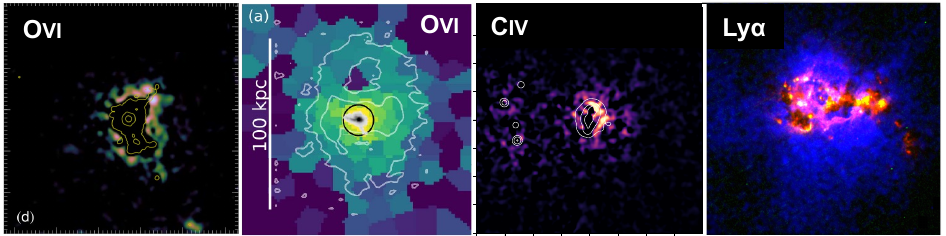}
    \caption{Composite results from SBC filter synthesis. Leftmost: two \oVI\ images of J1156+5008 (a $z=0.24$ starburst) and the `Makani' galaxy from \citet{Hayes.2016} and \citet{Ha.2025}, respectively.  Third: \cIV\ image of the $z=0.013$ starburst galaxy J1044+0353, synthesized from  data in GO\,16209. Rightmost: a \lya\ image (in blue) of the $z\sim 0$ starburst ESO\,338-IG04 \citep{Ostlin.2009}. }
    \label{fig:lineimages}
\end{figure}

\subsection{Ultraviolet Imaging of Emission Lines}\label{sect:uvlineimaging}

A  unique and irreplaceable capability of HST is its ability to image emission lines in the far ultraviolet. At $\lambda <1650$\,\AA, narrowband synthesis techniques with the Solar Blind Channel (SBC) of the \emph{Advanced Camera for Surveys} (ACS) can be used to create well defined bandpasses that can isolate line emission from continuum radiation \citep[][see Figure~\ref{fig:bandpass}]{Hayes.2009}. This has been verified for \hI\ \lya\ in $\simeq 100$ galaxies \citep{Ostlin.2009, Melinder.2023, Runnholm.2023, LeReste.2025.LaCOS},
and nebular \cIV\ imaging of NGC 2366 yields the first direct confirmation of radiative cooling from starburst feedback  \citep{Oey.2023}.
\emph{Any} UV feature can be imaged in this way provided its contrast is high enough, and the target galaxy lies at the correct redshift.  The STIS/\cIII\ dedicated narrowband filter at $\lambda_\mathrm{c}\simeq 1900$\,\AA\ provides further line imaging capabilities at longer wavelength \citep[e.g.][and see submitted White Paper by M. Mingozzi]{Micheva.2020}, and narrow quad filters in WFC3/UVIS (e.g. FQ232N, FQ243) can be used at higher redshifts.  With these available bandpasses, many combinations of lines could be imaged by a hypothetical study, ranging from relatively low-ionization combinations (\lya+\cIV+\cIII) at $z\sim 0$ to higher ionization states (\neVIII+\oVI) at $z\sim0.8$. Possible combinations are listed in Table~\ref{tab:filter_redshifts}, and example results are shown in Figure~\ref{fig:lineimages}. 

\begin{figure}
    \centering
    \includegraphics[width=1\linewidth]{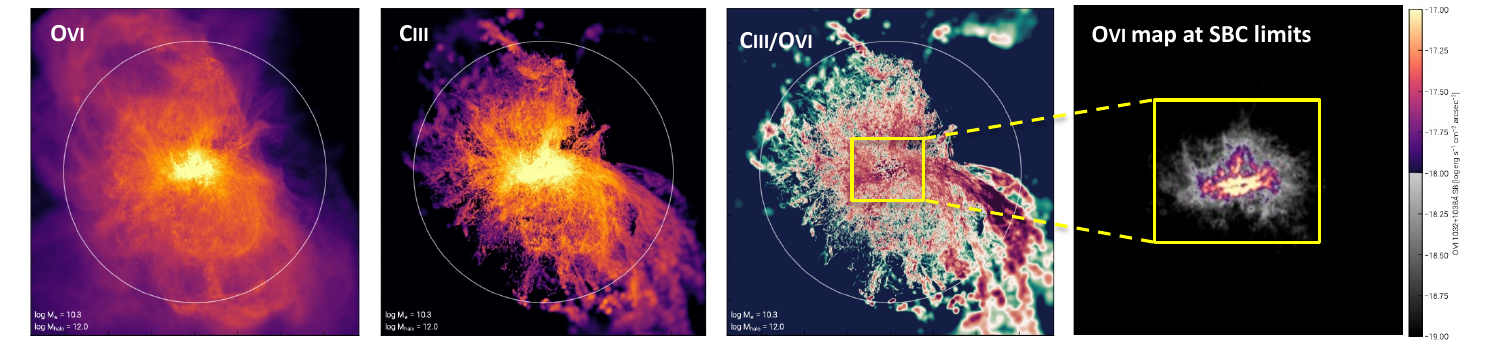}
    \caption{Simulations of expected \oVI\ and \cIII\ surface brightness using the GIBLE simulations \citep{Ramesh.2024}. Both peak above $10^{-18}$\,\ergseccmarcsec\  which is observable by SBC (Fig~\ref{fig:lineimages}).  The CIII/OVI ratio diagram, which varies by 2 orders of magnitude, includes diagnostic information of temperature and discriminates between volume-filling material and gas mixing in narrow sheets \citep{Chen.2023,Chen.2025}. The right panel shows the \oVI\ emission map with the color bar cut at the HST detection limit of \citet{Hayes.2016}, with the SBC field-of-view overlaid.}
    \label{fig:sims}
\end{figure}

While few examples exist to date, imaging observations with ACS/SBC have revealed the glow of warm \oVI\ ions out to tens of kpc (Figure~\ref{fig:lineimages}). In the $z\sim 0.4$ starburst `Makani', these observations trace emission from starburst-driven winds out to $\sim 100$\,kpc \citep{Ha.2025}. However, it is the combination of imaging and spectroscopic data that is truly transformative.  Spectroscopic absorption line measurements determine the gas column density $N_\mathrm{ion}$, which is the product of $n_\mathrm{ion} L$, where $n_\mathrm{ion}$ is the volume density and $L$ the length of the absorbing column.  In contrast,  imaging measures surface brightness, which is proportional to the product of the emission measure $n_\mathrm{ion} n_\mathrm{e} L$.  The well-known hurdle in interpreting either observation is that neither density nor column length can be inferred, which outlines the need for model-driven inference to estimate these quantities. However if both imaging and spectroscopy are obtained for the same object (either down-the-barrel or toward a background sightline), then the different dependence on $n$ allows observers to crack this degeneracy and infer volume densities and cloud sizes.  This is transformative in our understanding of the tenuous media around galaxies. 

Where both imaging and spectroscopy were obtained for J1156+5008 \citep[][Fig~\ref{fig:lineimages} left]{Hayes.2016}, the authors could infer the \oVI\ bearing gas to be confined to relatively narrow clouds ($\simeq 10$\,pc in length, filling $\sim 10^{-3}$ of the volume) of intermediate density ($\simeq 0.1$\,\percc).  These densities are very low compared to typical ISM densities ($\sim1-10^3$\,\percc), but very high compared to those inferred for the diffuse CGM ($\sim 10^{-3}$\,\percc) of normal galaxies \citep[e.g.,][]{Werk.2014,Prochaska.2017,Pointon.2019,Zahedy.2021}.  Given the small cloud sizes it is likely the observations trace narrow mixing layers that arise where the hot starburst-driven outflow interacts with cooler ambient gas in the CGM.  This is further supported by inferences from the same data, that show both the cooling times and sound-crossing times of the clouds must be exceedingly short ($\sim 10^5 - 10^6$ years) compared to typical starburst lifetimes and the times needed to lift gas from central regions to the radial distances at which emission is observed. This gas phase, therefore, must be constantly replenished. 

Unfortunately, such observations are both very rare in the HST archive/literature and expensive to execute.  They cost $\simeq 20$ orbits per target for imaging alone, with more required for the absorption spectroscopy.  Only 6 galaxies have \oVI\ imaging observations obtained, with two publications -- consequently we have no idea how energy is radiated as a function of physical quantities (e.g. SFR), and we have no understanding of the overarching implications for gas cooling, feedback efficiency, or galaxy formation.  Observations have been obtained in \cIV 1550 for three galaxies (Figure~\ref{fig:lineimages}), but possible features like \cIII\ 977\AA\ and \neVIII\ 775\AA, which trace gas near $10^5$ and $10^6$\,K respectively and could have similar or very different origins, have never been targeted with HST. Ratios of these features encode the origin of this intermediate temperature material (see Figure~\ref{fig:sims} and \citealt{Chen.2025} for details).  These observations, which have the power to answer all the questions outlined in Box 1, represent a potential legacy of the telescope that is not close to fully exploited.

\section{Large Scale Programs and Preparation for Habitable Worlds Observatory}\label{sect:hwoprep}

The delivery of a substantial suite of UV emission line images is a potential untapped legacy of HST's UV capabilities.  No other facilities, either in existence or planned, have the capabilities to deliver this data before HWO.  For example, the upcoming UVEX satellite has neither the narrowband filters to undertake the emission line imaging, nor the high resolution spectrographs to do the absorption line work (UVEX carries only a $R\simeq 1000$ slit spectrograph and the detector background prevents deep emission line imaging of faint diffuse targets). Aspera \citep{Vargas.2025} will map \oVI\ emission in three low-$z$ galaxies, and will not provide statistical samples nor have the resolution to undertake absorption line work. Explorer class missions can obtain such observations, but significant samples require the large primary mirror of Hubble.   \textbf{A boost to HST is the only way to get there within 20 years.} 

The objective of such an observing campaign must be to provide combined imaging and spectroscopic observations of statistically meaningful samples of galaxies spanning ranges of interesting/relevant parameters \{stellar mass, SFR, etc.\}.  Existing programs to image \oVI\ have used 16--20 orbits of SBC imaging for one line, and will require at least 5 orbits of COS spectroscopy to measure the absorption lines down-the-barrel.  Programs aiming to obtain images of two lines would then require $\simeq 40$ orbits/target, so a statistically significant sample of 25 galaxies that would be needed to explore scaling relations would arrive at $\sim 1000$ orbits.  This is larger than any single HST campaign ever executed, and needs a dedicated `XXL' investment akin to the ULLYSES Director's Discretionary Time program. 

HWO is likely to fly with both an ultraviolet integral field spectrograph (IFS) and a MOS. Instrument parameters are currently not determined, but the large array of Science Case Development Documents emphasized the need for spectroscopy at 1000-1500\AA, as did the 2020 Decadal Survey.  However, under the assumption this spectral imager will come close to Nyquist sampling the PSF, the spaxel areas are likely to be very small. Hence the field-of-view will also be small, making the mapping of large areas an extremely time consuming task.  Thus, large-FoV narrowband imaging delivered by HST could provide important guidance on targeting.  The HST data delivered by a program such as this could guide commissioning observations and early science explorations, optimizing the early use of HWO. 

\section{Necessary Observing Modes}

\textbf{Diffuse gas imaging at low surface brightness.} UV emission lines can be imaged using the nested long-pass filters of \textbf{ACS/SBC}. These share the same red wings but have different cut-on wavelengths; subtraction of adjacent pairs releases narrow, well-defined bandpasses that can be used to isolate emission lines \citep{Hayes.2009}. Thus, any emission line at $\lambda<1700$\,\AA\ can be imaged for galaxies at the specific redshifts. Figure~\ref{fig:bandpass} shows the redshift ranges at which specific lines can be imaged. All the LP filters of SBC are needed to capture specific line combinations. STIS hosts a narrowband filter to target \cIII\,1909 in the restframe, which captures other emission lines at higher $z$ (e.g. \lya\ at $z\simeq 0.58$). 

\textbf{Column density \& kinematics}. Mid-resolution gratings with \textbf{COS} have sufficient sensitivity and resolution to expose the stellar (or AGN) continuum.  The \textbf{G130M and G160M} elements are needed to measure the column densities and kinematics of the absorption lines (\cIV, \oVI, \neVIII, others). It is the combination of these column densities with the surface brightness measured in emission (ACS/SBC) that enable the derivation of volume densities, flow rates, cloud sizes, filling factors and all of the timescales relevant for gas cooling (see Section~\ref{sect:uvlineimaging}). 

\textbf{The ISM \& massive stars.}  Ultimately we need to connect the conditions in the CGM and galaxy winds/outflows to the properties of the stellar population.  This must include current generations of massive stars, as well as those forming some 100\,Myr in the past that could have launched winds in previous bursts.  Stellar conditions, including more extended star formation histories, can readily be inferred by fitting population synthesis modeling to the UV continuum \citep[e.g.][]{ficus.2025}.  The optimal observations would be spatially resolved such that (a.) the stellar conditions can be directly connected to the gas, and (b.) very recent episodes of massive star formation do not completely outshine the more evolved stars that formed $\simeq100$\,Myr in the past. The high spatial resolution spectroscopy needs \textbf{STIS MAMA slit modes}, possibly implementing slit-stepped observations, which could also map higher surface brightness emission (e.g., \cIV) at scales that cannot be delivered by COS.

\end{document}